\documentclass[aps,twocolumn,showpacs]{revtex4}
\usepackage{amsmath}
\usepackage{epsfig}

\begin{document}

\title{Hydrogen in a cavity}

\newcommand*{\NJNU}{Department of Physics, Nanjing Normal University, Nanjing, Jiangsu 210097, China}\affiliation{\NJNU}
\newcommand*{\NJU}{Department of Physics, Nanjing University, Nanjing, Jiangsu 210093, China}\affiliation{\NJU}

\author{Jialun Ping}\email{jlping@njnu.edu.cn}\affiliation{\NJNU}
\author{Hongshi Zong}\email{zonghs@nju.edu.cn}\affiliation{\NJU}

\begin{abstract}
The system of a proton and an electron in an inert and impenetrable spherical cavity
is studied by solving Schr\"{o}dinger equation with the correct boundary conditions.
The differential equation of a hydrogen atom in a cavity
is derived.
The numerical results are obtained with the help a power and efficient few-body method,
Gaussian Expansion Method. The results show that the correct implantation of the boundary
condition is crucial for the energy spectrum of hydrogen in a small cavity.
\end{abstract}


\maketitle

\section{\label{sec:introduction}Introduction}

The study of confined quantum system is an interesting topic recently~\cite{PR271}. With the advance of
technique, a number of confined quantum systems can constructed. For example, the well known
confined quantum systems are quantum wells, quantum wires and quantum dots~\cite{CQS}. The study
of the confined quantum system is helpful to understand the various properties of
nano-structures~\cite{nano}.

The simplest confined quantum system is that a hydrogen atom confined in a spherical cavity.
It was first investigated by Michels {\em et al.} about 80 years ago~\cite{first}, followed by
Sommerfeld and Welker~\cite{SW}. Since then the problems concerning confined atoms have been
studied by many authors~\cite{atom}. Various methods are introduced to solve the problem.
Perturbation methods~\cite{Hull}, variational methods~\cite{vari}, phase integral method~\cite{FYF},
etc.
In the previous work, people always assume that the proton in the hydrogen is fixed in the cavity,
because of the large mass of proton. This assumption is reasonable in the free space,
the two-body problem can be reduced to one-body problem by introducing the center-of-mass motion
and relative motion coordinates. The boundary condition is applied to the relative motion
coordinates of electron. However, the boundary condition should be applied to proton and electron
separately. In this case, the two body Schr\"{o}dinger equation can no longer be divided into the
center-of-mass and relative motion, and the situation becomes more complex. To develop a new method
to solve the problem of a hydrogen atom confined in an inert and
impenetrable spherical cavity is the goal of the present work. As a preliminary work, the angular
momentum is set to 0 and only the first three radial states are constructed.

In the present work, the motion of the proton is taken into account. The boundary condition is
applied both to the electron and proton motion. The problem is solved numerically with the help of
modified gaussian expansion~\cite{GEM}. The method is explained in the next section. The numerical results are
presented in the Sec. III. A brief summary is given in the last section.

\section{Method}

The proton-electron system in an impenetrable spherical cavity with radius $r_0$ is shown in Fig. 1.
The non-relativistic Hamiltonian of the system is (in atomic unit)
\begin{equation}
H^{ep} = -\frac{\nabla^2_1}{2}-\frac{1}{\bar{m}_p}\frac{\nabla^2_2}{2}-\frac{1}{r_{12}}+V(r_1)+V(r_2),
\end{equation}
where $\bar{m}_p$ is the mass of proton, $r_{12}$ is the distance between electron and proton.
The impenetrable spherical cavity is represented as
\begin{equation}
 V(r_{i}) =  \left\{ \begin{array}{ll}
 0 &
 \qquad r_{i} < r_0 \\
\infty  & \qquad
 r_{i} > r_0 \\
\end{array} \right.
\end{equation}

\begin{figure}[htb]
\begin{center}
\epsfxsize=1.5in \epsfbox{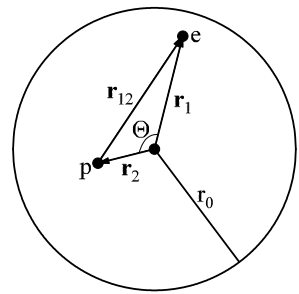} \vspace{-0.1in}

\caption{A hydrogen in a cavity.}
\end{center}
\end{figure}

The Schr\"{o}dinger equation to be solved is
\begin{eqnarray}
 & & \left[ -\frac{\nabla^2_1}{2}-\frac{1}{\bar{m}_p}\frac{\nabla^2_2}{2}-\frac{1}{r_{12}} \right]
 \Psi^{ep}(\mathbf{r}_1,\mathbf{r}_2) = E^{ep} \Psi^{ep}(\mathbf{r}_1,\mathbf{r}_2), \nonumber \\
 & & ~~~~~~~~~~~~~~~~~~~~~~~~~~~~~~~~~~~~~~~\mbox{for } r_1,r_2 < r_0
\end{eqnarray}
with boundary conditions
\begin{equation}
 \Psi^{ep}(r_1 \geq r_0,\mathbf{r}_2) = \Psi^{ep}(\mathbf{r}_1,r_2\geq r_0) =0
\end{equation}

Because of the cavity, the spatial translational invariance of the system is violated.
To separate the motion of the system into center-of-mass motion and relative motion by introducing
the Jacobi coordinates is meaningless. However, to remove the center-of-mass kinetic energy
is necessary for studying the hydrogen atom in a cavity. So Hamiltonian of a hydrogen atom
in a cavity is modified as:
\begin{eqnarray}
H^{H} & = & -\frac{\nabla^2_1}{2}-\frac{1}{\bar{m}_p}\frac{\nabla^2_2}{2}
+\frac{1}{2(1+\bar{m}_p)}\left(\boldsymbol{\nabla}_1+\boldsymbol{\nabla}_2\right)^2 -\frac{1}{r_{12}}\nonumber \\
& & +V(r_1)+V(r_2).
\end{eqnarray}

The Schr\"{o}dinger equation for a hydrogen in a cavity is
\begin{eqnarray}
 & & \left[ -\frac{\nabla^2_1}{2}-\frac{1}{\bar{m}_p}\frac{\nabla^2_2}{2}
 +\frac{1}{2(1+\bar{m}_p)}\left(\boldsymbol{\nabla}_1+\boldsymbol{\nabla}_2\right)^2-\frac{1}{r_{12}} \right]
 \nonumber \\
 & &  \Psi^{H}(\mathbf{r}_1,\mathbf{r}_2)= E^{H} \Psi^{H}(\mathbf{r}_1,\mathbf{r}_2), \quad  ~~~~\mbox{for } r_1,r_2 < r_0
\end{eqnarray}
with boundary conditions
\begin{equation}
 \Psi^{H}(r_1 \geq r_0,\mathbf{r}_2) = \Psi^{H}(\mathbf{r}_1,r_2\geq r_0) =0
\end{equation}

The standard procedure to solve the equation for hydrogen in a free space is to introduce the relative motion
and center-of-mass coordinates, $\mathbf{r}$ and $\mathbf{R}$. However, this procedure does not work for the
hydrogen in a cavity because of the boundary conditions, we cannot setup the proper boundary condition for
the relative motion and the center-of-mass motion. We have to solve the equation for the
hydrogen in a cavity using  independent coordinates $\mathbf{r}_1$ and $\mathbf{r}_2$.

Due to the spherical symmetry, the wavefunction of the hydrogen in a cavity $\Psi^{H}(\mathbf{r}_1,\mathbf{r}_2)$
can be written as $\Psi^{H}(r_1,r_2,x=\cos\Theta)$ (see Fig. 1). Using $r_1,r_2,x$, the Hamiltonian can be written as
(for $r_1,r_2 < r_0$)
\begin{widetext}
\begin{eqnarray}
H^H & = & -\frac{1}{\sqrt{r_1^2+r_2^2-2r_1r_2x}}
 -\frac{\bar{m}_p}{2(1+\bar{m}_p)}\left\{\frac{1}{r_1^2}\frac{\partial}{\partial r_1}
 \left( r_1^2\frac{\partial}{\partial r_1}\right) +\frac{1}{r_1^2}\left[ (1-x^2)\frac{\partial^2}{\partial x^2}-2x
  \frac{\partial}{\partial x}\right]\right\} \nonumber \\
  & & -\frac{1}{2\bar{m}_p(1+\bar{m}_p)}\left\{\frac{1}{r_2^2}\frac{\partial}{\partial r_2}
 \left( r_2^2\frac{\partial}{\partial r_2}\right) +\frac{1}{r_2^2}\left[ (1-x^2)\frac{\partial^2}{\partial x^2}-2x
  \frac{\partial}{\partial x}\right]\right\} \nonumber \\
  & & +\frac{1}{(1+\bar{m}_p)}\left\{  x\frac{\partial^2}{\partial r_1\partial r_2}+\frac{1-x^2}{r_2}\frac{\partial^2}{\partial r_1\partial x}
 +\frac{1-x^2}{r_1}\frac{\partial^2}{\partial r_2\partial x} -\frac{x(1-x^2)}{r_1r_2}\frac{\partial^2}{\partial x^2}
 +\frac{(1+x^2)}{r_1r_2}\frac{\partial}{\partial x}\right\} \nonumber \\
\end{eqnarray}
\end{widetext}

To find the analytic solution of $\Psi^{H}(r_1,r_2,x)$ is too difficult to be done. So the numerical method is
employed. For the sake of simplicity in this work, only the $L=0$ states are considered. Here the Gaussian expansion method, a powerful
method for few-body system with high precision is used~\cite{GEM}. The wavefunction $\Psi^{H}(r_1,r_2,x)$ is expanded as
\begin{eqnarray}
\Psi^{H}(r_1,r_2,x) & = & \frac{\sin\frac{\pi r_1}{r_0}\sin\frac{\pi r_2}{r_0}}{r_1r_2}\sum_{n=1}^{n_{max}} c_{n}
 e^{-\nu_n r_{12}^2}.
\end{eqnarray}
The gaussian size parameters are taken in geometric progression
\begin{equation}
\nu_n=\frac{1}{b_n^2},~~~~b_n=b_1 a^{n-1},~~~~a=\left(\frac{b_{n_{max}}}{b_1}\right)^{\frac{1}{n-1}}.
\end{equation}
For the small cavity, $r_0 \leq 1$, $n_{max}$ is 10 at most. For the large cavity, $r_0 \geq 20$,
$n_{max}=30$ is enough for getting the converged results.

\section{results}
\begin{table*}[ht]
\caption{\label{results}The eigen-energies of hydrogen in a cavity and average distance between the electron
and the proton.}
\begin{tabular}{ccccccccccc}  \hline\hline
 $r_0$ &       & \multicolumn{6}{c}{GEM} & \multicolumn{3}{c}{Ref.\cite{JPC92}} \\   \hline
       &       &  \multicolumn{3}{c}{$H^H$} &\multicolumn{3}{c}{$H^0$} &   & &  \\   \hline
       &       &   $1S$  &  $2S$   &  $3S$   &   $1S$  &  $2S$   &  $3S$   &   $1S$  &  $2S$   &  $3S$   \\ \hline
 0.10  & ~~$E$~~   &  475.22 & 1103.8  & ~~1975.0 ~~ &  469.00 & 1942.7  & 4406.8  &  468.99 & 1942.7 & 4406.1  \\
       & $d$   &  0.0692 & 0.0848  & 0.0907  &  0.0497 & 0.0500  & 0.0525  &         &        &         \\  \hline
 0.25  & $E$   &  71.634 & 172.08  & 311.24  &  69.096 & 303.32  & 696.54  &  69.094 & 303.31 & 696.51  \\
       & $d$   &  0.1702 & 0.2117  & 0.2268  &  0.1232 & 0.1251  & 0.1252  &         &        &         \\  \hline
 0.50  & $E$   &  16.014 & 41.109  & 75.818  &  14.749 & 72.674  & 170.59  &  14.748 & 72.672  & 170.59  \\
       & $d$   &  0.3306 & 0.4227  & 0.4542  &  0.2425 & 0.2506  & 0.2508  &         &         &         \\  \hline
 1.00  & $E$   &  2.9914 & 9.3042  & 17.948  &  2.3741 & 16.571  & 40.864  &  2.3740 & 16.570  & 40.863  \\
       & $d$   &  0.6192 & 0.8440  & 0.9111  &  0.4683 & 0.5033  & 0.5036  &         &         &         \\  \hline
 2.00  & $E$   &  0.1591 & 1.8260  & 3.9770  & $-0.12497$ & 3.3276  & 9.3143  & $-0.12500$ & 3.3275  & 9.3142  \\
       & $d$   &  1.0543 & 1.6879  & 1.8359  &  0.8594 & 1.0220  & 1.0169  &         &         &        \\  \hline
 4.00  & $E$   & $-0.3772$ & $ 0.2035$ & $ 0.7389$ & $-0.48324$ & $ 0.42026$ & $ 1.8727$ & $-0.48327$ & $ 0.42024$ & $ 1.8727$ \\
       & $d$   &  1.4185 & 3.2982  & 3.7356  &  1.3417 & 2.1462 & 2.0863   &         &         &         \\  \hline
 6.0   & $E$   & $-0.4498$ & $-0.0220$ & $ 0.2187$ & $-0.49926$ & $0.01273$ & $ 0.63175$ & $-0.49927$ & $0.01273$ & $ 0.63174$ \\
       & $d$   &  1.4843 & 4.5154  & 5.6291  &  1.4810 & 3.3080 & 3.2209   &         &         &         \\  \hline
 8.0   & $E$   & $-0.4722$ & $-0.0802$ & $0.0640$ & $-0.49996$ & $-0.08473$ & $0.24650$ & $-0.49998$ & $-0.08474$ & $0.24649$ \\
       & $d$   &  1.497 & 5.259  & 7.425  &  1.4987 & 4.2946 & 4.4083   &         &         &         \\  \hline
10.0   & $E$   & $-0.4818$ & $-0.1008$ & ~$0.0042$~ & ~$-0.49998$~ & $-0.11280$ & $0.09142$ & $-0.499999$ & $-0.112806$ & $0.091422$ \\
       & $d$   &  1.5010 & 5.6494  & 9.0361  &  1.5000 & 5.0258  & 5.6318 &          &         &         \\  \hline
25.0   & $E$   & $-0.4955$ & $-0.1219$ & $-0.0519$ & $-0.49999$ & $-0.12500$ & ~$-0.054592$~ & $-0.500000$ & $-0.125000$ & $-0.054592$ \\
       & $d$   &  1.5056 & 6.0045  & 13.304  &  1.5000 & 6.0000 & 12.707  &         &         &        \\ \hline
30.0   & $E$   & ~$-0.4975$~ & $-0.1230$ & $-0.0532$ & $-0.49999$ & $-0.12500$ & $-0.05542$ & $-$ & $-$ & $-$ \\
       & $d$   & 1.5019  & 6.0016  & 13.4137  &  1.5000 & 6.0000  & 13.307 &         &         &        \\ \hline
40.0   & $E$   &~$-0.4985$~ & $-0.1238$ & $-0.0544$ & $-0.50000$ & $-0.12500$ & $-0.05556$ & $-0.50000$ & $-0.12500$ & $-0.05556$ \\
       & $d$   &  1.5015 & 6.0034  & 13.481  &  1.5000 & 6.0000  & 13.496  &         &        &        \\
\hline\hline
\end{tabular}
\end{table*}

In order to check the precision of GEM, we first do a calculation of the hydrogen in the cavity
with proton fixed at the center of the cavity. In this case, the Hamiltonian is simplified to
\begin{equation}
H^{0} = -\frac{\nabla^2_1}{2} -\frac{1}{r_{1}} +V(r_1).
\end{equation}
The numerical results are obtained by using GEM and are shown in the columns with head $H^0$ of
Table \ref{results}, where only the eigen-energies and average distance $d$ between the electron
and the proton of first three radial states, $1S$, $2S$ and $3S$ are presented.
The energies agree with the previous results very well. The agreement shows that GEM is a effective method
with high precision for the confined quantum systems.

\begin{center}
\epsfxsize=3.8in \epsfbox{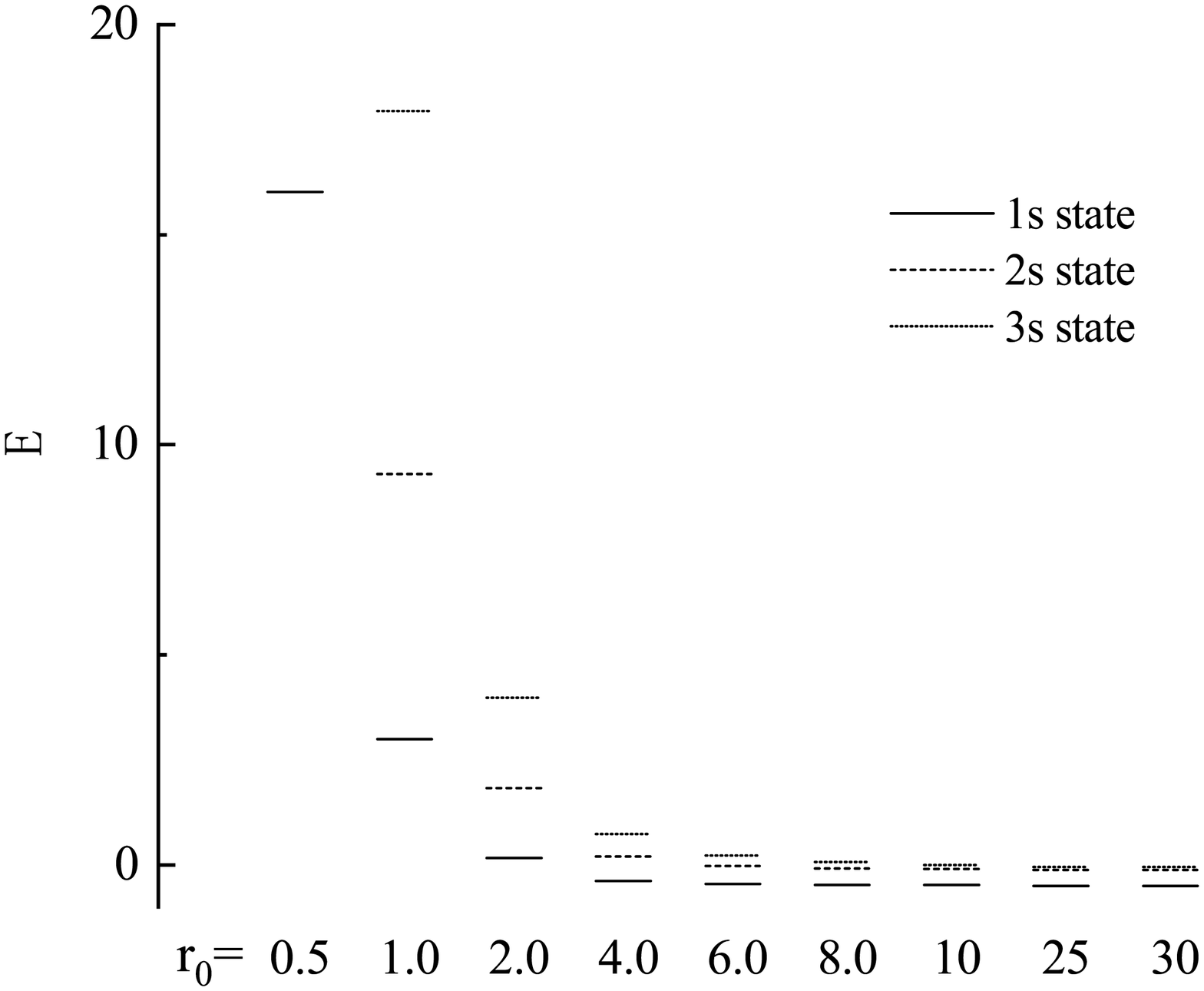} \vspace{-0.1in}

FIG. 2: The energy levels of a hydrogen in a cavity.
\end{center}

Considering the motion of proton and using correct boundary conditions, the eigen-energies and average
distance $d$ between the electron and the proton are also shown in Table \ref{results}.
Comparing with the results with $H^0$, there are significant differences, especially for the excited states.
For the ground state, our results are a little higher than the previous results.
For $r_0 =0.1$, there is a 1\% difference. With the increasing cavity radius, the difference decreases.
For the large cavity, $r_0 =50$, the difference will disappears. The remained small difference comes from
the reduced masses of electron is used in our calculation.
For the $3S$ state, our results deviate from the previous results rather large for $r_0\leq 10$.
We get a smaller energies, which is unusual, compared with the ground state. Further study is needed.

The energy levels of the system are shown in Fig. 2. We can see that for the small cavity, the boundary condition
has larger effects on the hydrogen, the energy difference between $3S$ and $2S$ is bigger that that between
$2S$ and $1S$ (for $r_0=1.0$, $E_{3S}-E_{2S}=8.634$, $E_{2S}-E_{1S}=6.3028$), this is the feature of a particle
moving in a spherical well. For the larger cavity, the Coulomb potential between electron and proton will
become dominant, the energy difference between $3S$ and $2S$ is much smaller that that between
$2S$ and $1S$ (for $r_0=25$, $E_{3S}-E_{2S}=0.07$, $E_{2S}-E_{1S}=0.3736$), the feature of a particle moving
in a Coulomb potential.

\begin{center}
\epsfxsize=3.5in \epsfbox{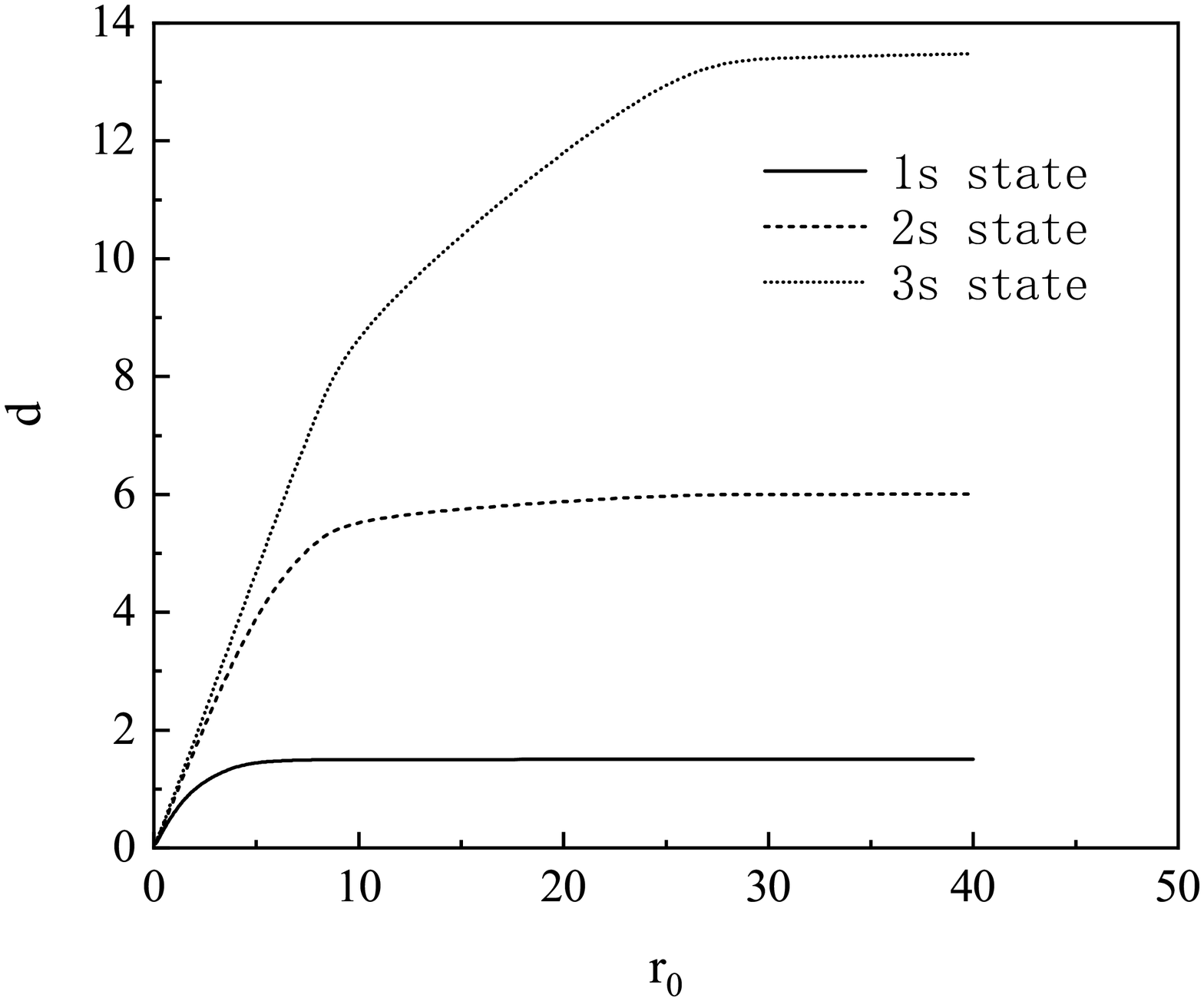} \vspace{-0.1in}

FIG. 3: The average distance between the electron and the proton of a hydrogen in a cavity.
\end{center}

For a small cavity, the binding energy of the system is positive, which means that the bound state is formed by the
cavity, not by Coulomb potential between the electron and the proton. We also calculate the critical radius of the
cavity for zero binding energy of the hydrogen, which is about 2.2363, and in this case the average distance between
the electron and the proton is around 1.1288.

Fig. 3 displays the variation of the average distance with the radius of the cavity. Clearly the cavity has much
stronger influence on the excited states, because the excited states spread more.

\section{summary}
By considering the motion of proton and using the correct boundary conditions for electron and proton,
the hydrogen in an inert and impenetrable spherical cavity are studied by solving Schr\"{o}dinger equation.
A powerful few-body method, GEM is employed to do a numerical calculation.
The results show that for a not too large cavity, our results are different from the previous ones
with fixed proton. So the correct mount of boundary conditions is important for the confined quantum
systems.

It is worth to mention that the center-of-mass motion of the system is removed before solving Schr\"{o}dinger
equation. Including the center-of-mass motion, solving Schr\"{o}dinger equation, then separating the
center-of-mass motion, is another story because of the violation of the translation invariance.

In the present work, only the first three radial states are considered. To generalized the calculation to
other states are straightforward, which is our next work.

\section*{Acknowledgment}
This work is supported partly by the National Science Foundation
of China under Contract Nos. 11775118 and 11535005.

\end{document}